\documentclass[12pt]{article}
\usepackage[utf8]{inputenc}
\usepackage{amsmath}
\usepackage{amsfonts}
\usepackage{authblk}
\usepackage{multirow}
\usepackage{float}
\usepackage{graphicx}
\usepackage{booktabs}
\usepackage[titletoc,toc,title]{appendix}
\usepackage{color}

\newcommand{\be}{\begin{equation}}
\newcommand{\ee}{\end{equation}}
\newcommand{\baln}{\begin{eqnarray}}
\newcommand{\ealn}{\end{eqnarray}}
\newcommand{\ben}{\begin{equation*}}
\newcommand{\een}{\end{equation*}}
\newcommand{\nn}{\nonumber}
\newcommand{\ta}{\tilde{a}}

\title{
Universal behaviour of generalized Causal set d'Alembertians in curved spacetime}

\author{Alessio Belenchia$^{1,2}$}
\affil{$^1$SISSA, via Bonomea 265, Trieste, Italy}
\affil{$^2$INFN sezione di Trieste, via Valerio 2, Trieste, Italy}
\date{\today}

\begin{document}

\maketitle

\begin{abstract}
Causal set non-local wave operators allow both for the definition of an action for Causal set theory and the study of deviations from local physics that may have interesting phenomenological consequences. It was previously shown that, in all dimensions, the (unique) minimal discrete operators give averaged continuum non-local operators that reduce to $\Box-R/2$ in the local limit. Recently, dropping the constraint of minimality, it was shown that there exist an infinite number of discrete operators satisfying basic physical requirements and with the right local limit in flat spacetime. In this work, we consider this entire class of Generalized Causal set d'Alembertins in curved spacetimes and extend to them the result about the universality of the $-R/2$ factor. Finally, we comment on the relation of this result to the Einstein Equivalence principle. 

\end{abstract}

\vspace{2pc}

\section{Introduction}\label{Intro}
Causal set (CS) theory is a proposal for a Quantum Gravity (QG) theory that assumes a discrete structure for spacetime (see~\cite{Surya:2011sf} and~\cite{Henson:2006kf} for reviews on CS theory, with emphasis on modern developments). The critical assumptions of the theory are discreteness and Lorentz Invariance. Maintaining (local) Lorentz Invariance while making spacetime fundamentally discrete comes with a price, causal sets are inherently non-local. This can be intuitively understood in a causal set approximating Minkowski spacetime where Lorentz Invariance and discreteness imply that, given a point, the nearest neighbours are all the points of the CS one Planck time away from the chosen one, assuming the discreteness scale to be the Planck one. It is clear that those points are infinite in number and distributed near the null cone up to infinity, making the CS an extremely non-local object. This non-locality plays a central role in determining the dynamics of matter over a fixed CS.

There are different ways for describing the propagation of (scalar) fields on causal sets~\cite{Johnston:2010su}. One way is through the introduction of discrete d'Alembertian operators~\cite{Sorkin:2007qi} defined on generic causal sets that do not require any embedding spacetime. These operators, once averaged over all sprinklings on flat spacetime, give rise to the standard wave operators in the local limit~\cite{Benincasa:2010ac,Glaser:2013sf,Dowker:2013vl,Aslanbeigi:2014tg}. It is in this sense that they are discrete versions of the standard d'Alembertian. Similarly, when averaged over all sprinklings on a given curved spacetime the operators reduce in the local limit to the covariant d'Alembertian plus a term proportional to the scalar curvature. In particular, they reduce to
$$\Box\phi(x)-\frac{1}{2}R(x)\phi(x)$$ in the local limit for all dimensions, i.e. the factor $-R/2$ is universal~\cite{Dowker:2013vl,Glaser:2013sf}. 

In this work we re-derive the result on the universality of the $-R/2$ factor and extend it to the whole family of \textit{Generalized Causal set d'Alembertians} (GCD)introduced in~\cite{Aslanbeigi:2014tg} (which contain the operators studied in~\cite{Dowker:2013vl,Glaser:2013sf} as special cases). In doing so, we bridge the gap between the spectral analysis --- difficult to generalized to curved spacetime --- used in~\cite{Aslanbeigi:2014tg} and the way in which causal set d'Alembertians are studied in curved spacetime~\cite{Belenchia:2015hca}.
Finally, we discuss the implications of this result in relation to the Einstein Equivalence Principle.  
\\
The paper is organized as follows. In section~\ref{II} we introduce both the GCD family of operators and the general geometric set-up with which we will work. Section~\ref{III} constitutes the proof of the universality of the $-R/2$ factor and in section~\ref{V} we conclude with a discussion of the result and future directions.


\section{General set-up}\label{II} 
In this section we introduce the family of GCD~\cite{Aslanbeigi:2014tg} and outline the general set-up used in the rest of the work. 

\subsection{Generalized Causal set d'Alembertians}\label{GCDu}
    
Given a causal set $\mathcal{C}$ and a scalar field $\phi:\mathcal{C}\rightarrow\mathbb{R}$ on it, let us consider the operators defined by\footnote{Here we follow the notation of~\cite{Aslanbeigi:2014tg}.} 
\be\label{opn}
(B^{(D)}_{\rho}\phi)(x)=\rho^{2/D}\left(a\,\phi(x)+\sum_{n=0}^{L_{max}}b_{n}\sum_{y\in I_{n}(x)}\phi(y)\right),
\ee
where $a,b_{n}$ are dimension dependent coefficients, $\rho=l^{-D}$, $l$ is the discreteness scale and $I_{n}(x)$ represents the set of past $n$-th neighbours\footnote{A point $y$ is an $n$-th past neighbour of $x$ if the cardinality of the set $Int(x,y)=\left\{z\in\mathcal{C}: x\prec z\prec y\right\}$ is equal to $n$.} of $x$. In the literature the first sum in eq.~\eqref{opn} is referred to as sum over \textit{layers}, where each $I_{n}$ is a layer. The operators in eq.~\eqref{opn} are derived under the following physical assumptions~\cite{Aslanbeigi:2014tg}
\begin{enumerate}
\item \textbf{Linearity}: the result of the action of the operator on a scalar field at an element of the causal set should be a linear combination of the values of the fields at other elements
\item\textbf{Retardedness}: the result of the action of the operator on a scalar field at an element of the causal set should depend on the values of the field in the past of that element
\item\textbf{Label invariance}: the operator should be invariant under the relabellings of the causal set elements
\item\textbf{Neighbourly democracy}: all $n$-th neighbours of $x$ should contribute to $(B^{D}_{\rho}\phi)(x)$ with the same coupling
\end{enumerate}
The operators in eq.~\eqref{opn} are well defined for a general causal set $\mathcal{C}$ but are particularly relevant for causal sets that well approximate continuum spacetimes~\cite{Belenchia:2015hca}. Indeed, given a generic spacetime\footnote{Actually we have to consider spacetimes that satisfy some causality conditions. The minimal requirement for the causal set itself to be meaningful is future and past distinguishability.} $(\mathcal{M},g)$, the average of the discrete operators over all Poisson sprinklings of $\mathcal{M}$ lead to a continuum operator given by\footnote{For further details see~\cite{Surya:2011sf} and references therein.}  
\begin{align}\label{cont}
\mathbb{E}(B^{(D)}_{\rho}\phi)(x)&=\rho^{2/D}a\,\phi(x)\\ \nonumber
&+\rho^{(2+D)/D}\sum_{n=0}^{L_{max}}\frac{b_{n}}{n!}\int_{J^{-}(x)}\sqrt{-g}e^{-\rho V(x,y)}[\rho V(x,y)]^{n}\phi(y) d^{D}y,
\end{align}
where $\mathbb{E}$ stands for average over sprinklings, $J^{-}(x)$ is the causal past of $x$ and $V(x,y)$ is the spacetime volume of the causal interval between $x$ and $y$.
It was shown in~\cite{Sorkin:2007qi, Dowker:2013vl} that, for particular choices of coefficients $\left\{a,b_{n}\right\}$, the operators in eq.~\eqref{cont} reduce to the standard wave operator in flat spacetime in the local limit, i.e. for $\rho\rightarrow\infty$. Once $\left\{a,b_{n}\right\}$ are identified, these operators can also be studied for a generic curved spacetime~\cite{Dowker:2013vl,Glaser:2013sf,Belenchia:2015hca}. 

GCD were studied in flat spacetime in~\cite{Aslanbeigi:2014tg}, where relations defining the coefficients $\left\{a,b_{n}\right\}$ were found via a spectral analysis. In the following subsections we give an overview of these relations that are fundamental to this work.

\subsubsection{Even dimensions}\label{even}
The coefficients $\left\{a,b_{n}\right\}$ in even dimensions, defining $D=2N+2$ with $N=0,1,\dots$, are determined by the following equations
\begin{subequations}
\begin{align}
\sum_{n=0}^{L_{max}}\frac{b_{n}}{n!}\Gamma(n+\frac{k+1}{N+1})&=0,\;\; k=0,1,\dots, N+1 \label{eq:even1} \\ 
a+\frac{2(-1)^{N+1}\pi^{N}}{N!D^2 C_{D}}\sum_{n=0}^{L_{max}}b_{n}\psi(n+1)&= 0 \label{eq:even2}\\ 
\sum_{n=0}^{L_{max}}\frac{b_{n}}{n!}\Gamma(n+\frac{N+2}{N+1})\psi(n+\frac{N+2}{N+1})&=\frac{2(-1)^N (N+1)!}{\pi^N}D^2 C_{D}^{\frac{N+2}{N+1}}, \label{eq:even3}
\end{align}
\end{subequations}
where $C_{D}=\frac{(\pi/4)^{\frac{D-1}{2}}}{D\Gamma\left(\frac{D+1}{2}\right)}$ (actually this definition is also valid in odd dimensions), $\Gamma$ is the Gamma function and $\psi$ stems for the Digamma function.

\subsubsection{Odd dimensions}\label{odd}
In odd dimensions, defining $D=2N+1$ with $N=0,1,\dots$, the equations are 
\begin{subequations}
\begin{align}
\sum_{n=0}^{L_{max}}\frac{b_{n}}{n!}\Gamma(n+\frac{2k+2}{2N+1})&=0,\;\; k=0,1,\dots, N \label{eq:odd1}\\
a+\frac{(-1)^{N}\pi^{N+\frac{1}{2}}}{D C_{D}\Gamma(N+\frac{1}{2})}\sum_{n=0}^{L_{max}}b_{n}&= 0 \label{eq:odd2}\\
\sum_{n=0}^{L_{max}}\frac{b_{n}}{n!}\Gamma(n+\frac{2N+3}{2N+1})&=\frac{4(-1)^{N-1} \Gamma(N+\frac{3}{2})}{\pi^{N+\frac{1}{2}}}D C_{D}^{\frac{2N+3}{2N+1}}. \label{eq:odd3}
\end{align}
\end{subequations}

Note that in even dimensions we have $N+4$ equations and in odd dimensions $N+3$. This means that in the minimal cases given by $L_{max}=\frac{D+2}{2}$ and $L_{max}=\frac{D+1}{2}$ in even and odd dimensions respectively, we have a unique solution corresponding to the minimal operators. In the non-minimal cases instead, the number of equations is less than the unknowns making the system under-determined therefore admitting an infinite number of solutions. 

\subsection{Geometrical set-up}\label{IIIno}
The continuum operators given by eq.~\eqref{cont} can be rewritten in the following form
\begin{align}\label{cont2}
\bar{B}^{(D)}_{\rho}\phi(x)\equiv\mathbb{E}(B^{(D)}_{\rho}\phi)(x)&=\rho^{2/D}a\,\phi(x)\\ \nonumber
&+\rho^{(2+D)/D}\hat{O}\underbrace{\int_{J^{-}(x)}\sqrt{-g}e^{-\rho V(x,y)}\phi(y) d^{D}y}_{I^{(D)}},
\end{align}
where the operator $\hat{O}$ is defined as
\be\label{ohat}
\boxed{
\hat{O}=\sum_{n=0}^{L_{max}}\frac{b_{n}}{n!}(-1)^{n}H_{n},
}
\ee
and $H_{n}\equiv\rho^{n}\partial^{n}/\partial\rho^{n}$. Note that $\hat{O}\rho^{n}\propto\rho^{n}$.

\begin{figure}[htb!]
 \centering
 \includegraphics[scale=0.3]{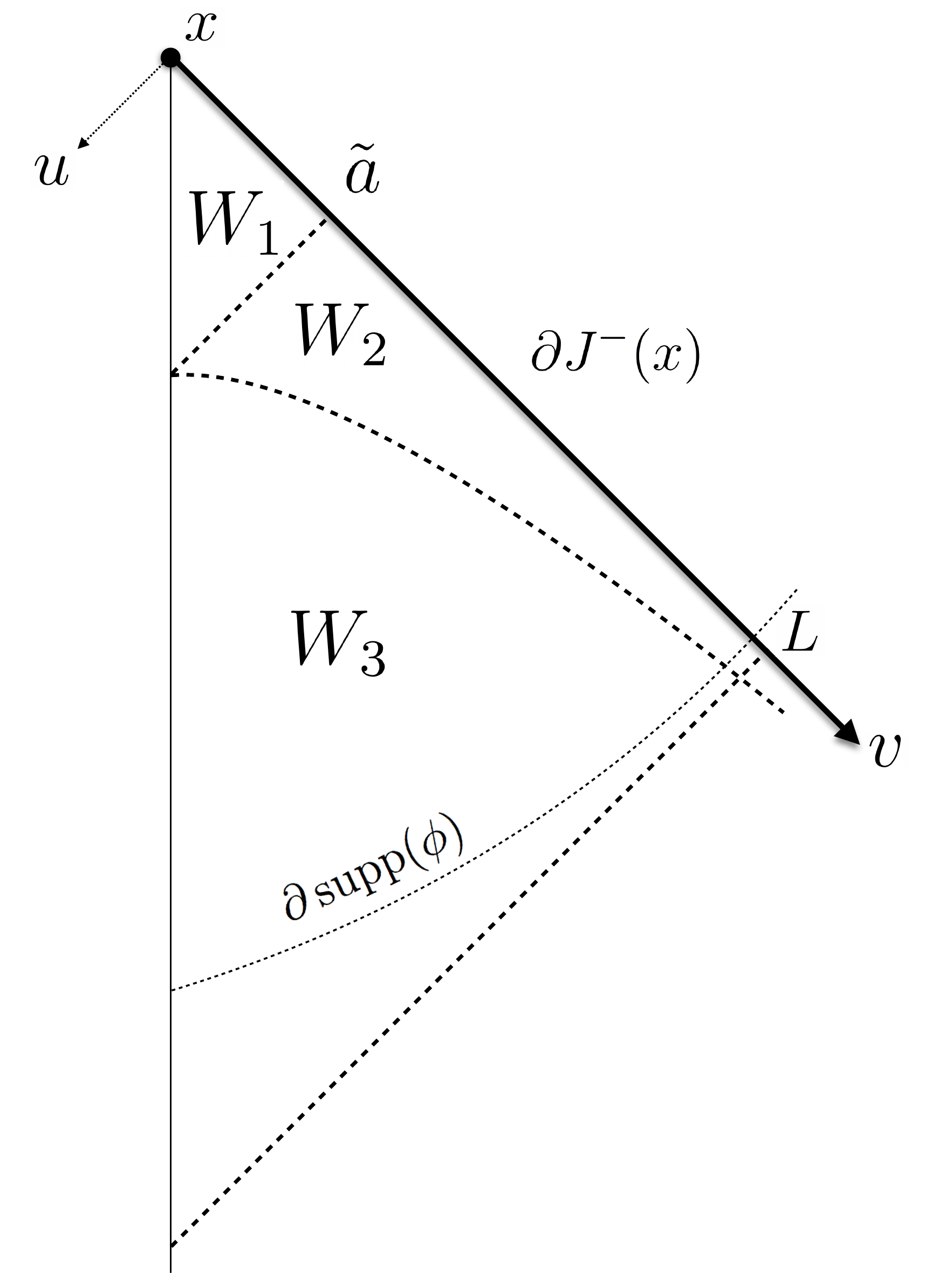}
 \caption{\label{nearregion}Graphical representation of the geometrical construction with $D-2$ dimensions suppressed. The dotted line represents the boundary of the intersection of the support of the field and the past light cone of the origin. Regions $W_{2}$ and $W_{3}$ are called \textit{down the light cone region} and \textit{far region} respectively and are not considered in this work. $W_{1}$ is the near region that we consider.}
\end{figure}

In the following we use the geometrical set-up of~\cite{Belenchia:2015hca} (see also~\cite{Dowker:2013vl}) and we assume that the scalar field has compact support of size smaller than the curvature radius and such that it varies slowly on scales of the order of the non-locality scale\footnote{Note that, even if the averaged operators converge to the local one in the local limit the fluctuations actually tend to increase. To tame this problem, a new scale $l_{k}$ was introduced in~\cite{Sorkin:2007qi}. In particular, new discrete operators have been found that give the same average as in eq.~\eqref{cont} with $\rho=1/l_{k}$. See~\cite{Aslanbeigi:2014tg} for further details.} $l_{k}$. In~\cite{Belenchia:2015hca} it was shown that the finite contributions to the local limit come from the so called \textit{near region} ($W_{1}$), i.e. the region of the past light cone of a chosen point $x$, that is a neighbourhood of the point itself, see Fig.\ref{nearregion}. 
Motivated by this result (see also discussion below) we focus on the integral $I^{(D)}$, in eq.~\eqref{cont2}, restricted to $W_{1}$ defined by
$$W_{1}:=\left\{y\in supp(\phi)\cap J^{-}(x): 0\leq u\leq v\leq \ta\right\},$$ where $\tilde{a}>0$ is chosen small enough for the expansions (introduced below) to be valid but such that the near region is much larger than the non-locality scale, i.e. $\rho\,\ta^{D}\gg 1$ (for more details on the geometrical constructions we refer the reader to~\cite{Belenchia:2015hca}).  In this region we use (past pointing) null Riemann normal coordinates $(u,v,\varphi_{1},...,\varphi_{D-2})$ defined by $u=(-y^{0}-r)/\sqrt{2}$, $v=(-y^{0}+r)\sqrt{2}$, where $r=\sqrt{\sum_{i=1}^{D-1}y^{i^{2}}}$ and $\left\{y^{\mu}\right\}$ are the RNC. 

The integral $I^{(D)}$ restricted to the near region in these coordinates takes the form
\be\label{interest}
I^{D}_{W_{1}}=\int_{0}^{\ta}\int_{0}^{v}du\int d\Omega_{D-2}\frac{(v-u)^{D-2}}{2^{(D-2)/2}}\sqrt{-g(y)}\phi(y)e^{-\rho V(y)}.
\ee
As already discussed in~\cite{Dowker:2013vl} (see Sec.3 therein) assuming the size of the compact support of the field to be smaller than the curvature radius implies that all the curvature corrections to the flat space integral will be small. 
We can assume then that eq.~\eqref{cont2} will converge to a local result as $\rho\rightarrow\infty$ as in the flat case. Thus, by dimensional arguments, the limit $\rho\rightarrow\infty$ of eq.~\eqref{cont2} will be a linear combination of the d'Alembertian and the Ricci scalar curvature.

What we will show in this work is that, when the aforementioned local limit exists, the limit will be $\Box_{g}-R/2$ for all GCD. 
The existence of a local limit is already assumed in earlier work~\cite{Glaser:2013sf,Dowker:2013vl}, where higher order terms were neglected due to this assumption and the dimensional arguments. In this work, even if we stick to the same assumption, we argue that the local limit always exists based on the results in~\cite{Benincasa:2010ac,Belenchia:2015hca} ( see appendix~\ref{app:extrat} and the following discussion).
It should be noted that our proof strictly holds true when the assumption that the compact support of the field is much smaller than the curvature scale is fulfilled. This implies that we do not have to consider regions $W_{2,3}$ at all, in the limit. 
However, this assumption could be relaxed by including regions $W_{2,3}$. Although such a proof goes beyond the scope of the present work, note that in~\cite{Belenchia:2015hca} the complete proof of the existence of a local limit in 4D curved spacetimes for the minimal operator is given considering all the terms and all the regions $W_{1,2,3}$. Since only the properties of the operator $\hat{O}$ are used to complete the proof we see no reason why the same construction can not be extended in all dimensions and to non-minimal operators. This lends strong support to the conjecture that the universality result we are going to prove extends to configurations like the one in Fig.~\ref{nearregion}, where the support of the field is not restricted to be in the near region.

In order to proceed, we expand the volume element and the volume of causal intervals around the origin up to order $R^{2}$ terms (see~\cite{Gibbons:2007nm, Dowker:2013vl}) and the field up to second derivative terms
\be\label{metricex}
\sqrt{-g}=1-\frac{1}{6}R_{\mu\nu}(0)y^{\mu}y^{\nu}+\mathcal{O}(R^{2})\equiv 1+\delta\sqrt{-g}+\mathcal{O}(R^{2}),
\ee
\begin{align}\label{volumeex}
V(y)&=V_{0}^{D}\left(1-\frac{D}{24(D+1)(D+2)}R(0)\tau^{2}+\frac{D}{24(D+1)}R_{\mu\nu}(0)y^{\mu}y^{\nu}+\mathcal{O}(R^{2})\right)\\ \nn
&\equiv V_{0}+\delta V+\mathcal{O}(R^{2}),
\end{align}
\be\label{fieldex}
\phi(y)=\phi(0)+y^{\mu}\phi_{,\mu}(0)+\frac{1}{2}y^{\nu}y^{\mu}\phi_{,\nu\mu}(0)+y^{\mu}y^{\nu}y^{\sigma}\Phi_{\mu\nu\sigma}(y),
\ee     
where $\tau^{2}=2 u v$, $V_{0}^{D}=C_{D}\tau^{D}=2^{D/2}C_{D}(uv)^{D/2}\equiv c_{D}(uv)^{D/2}$ is the volume of a causal interval in $D$-dimensional Minkowski spacetime between the origin and the point with Cartesian coordinates $\left\{y^{\mu}\right\}$ and $\Phi_{\mu\nu\sigma}(y)$ is a smooth function of $y$. 
Using these expansions we can then write eq.~\eqref{interest} as
\begin{align}\label{allt}
I^{(D)}&=\int_{0}^{\ta}dv\int_{0}^{v}du\frac{(v-u)^{D-2}}{2^{(D-2)/2}}\int d\Omega_{D-2} \\ \nonumber
& \left[(1+\delta\sqrt{-g})\cdot (\phi(0)+y^{\mu}\phi_{,\mu}(0)+\frac{1}{2}y^{\nu}y^{\mu}\phi_{,\nu\mu}(0))\cdot (1-\rho\delta V)\right.\\ \nn
&\left. +(1+\delta\sqrt{-g})\cdot (\phi(0)+y^{\mu}\phi_{,\mu}(0)+\frac{1}{2}y^{\nu}y^{\mu}\phi_{,\nu\mu}(0))\cdot \sum_{k=2}^{\infty}\frac{(-\rho\delta V)^{k}}{k!}\right]e^{-\rho V_{0}}.
\end{align}
where we neglected terms $\mathcal{O}(R^{2})$ and with more than two derivatives of the field coming from eqs.~\eqref{metricex},~\eqref{volumeex} and~\eqref{fieldex}. Since we are interested only in the local limit in the following we also neglect all terms $\mathcal{O}(R^{2})$ and $R\;\partial\phi $, $R\;\partial^{2}\phi$ in eq.~\eqref{allt}. Note that these terms are relevant when the nonlocality scale is not strictly vanishing (see~\cite{Belenchia:2015hca} for details in 4D with the minimal operator). We consider these these terms schematically in appendix~\ref{app:extrat}, and argue that they do not contribute in the local limit. 

Performing the integration over the spherical coordinates and using that $$\int d\Omega_{D-2} (y^{j})^{2}=\frac{1}{D-1}\int d\Omega_{D-2}=\omega_{D-1},$$ where $\omega_{D-1}$ is the volume of the Euclidean ball of radius one in $D-1$ dimensions, we arrive at 
\begin{align}\label{I}
I^{(D)}&=\int_{0}^{\ta}dv\int_{0}^{v}du\frac{(v-u)^{D-2}}{2^{(D-2)/2}} \\ \nonumber
& \left[(D-1)\omega_{D-1}\phi(0)+(D-1)\omega_{D-1}y^{0}\phi_{,0}(0)\right.\\ \nn
&\left.+\frac{(u+v)^{2}}{2}\omega_{D-1}(D-1)\left(\frac{1}{2}\phi_{,00}(0)-\frac{1}{6}R_{00}(0)\phi(0)-\rho\,\phi(0) c_{D}(uv)^{D/2}\frac{D}{24(D+1)}R_{00}(0)\right)\right.\\ \nn
&\left.+\frac{(v-u)^{2}}{2}\omega_{D-1}\left(\frac{1}{2}\phi_{,ii}(0)-\frac{1}{6}R_{ii}(0)\phi(0)-\rho\,\phi(0) c_{D}(uv)^{D/2}\frac{D}{24(D+1)}R_{ii}(0)\right)\right.\\ \nn
&\left.+ \omega_{D-1}(D-1)\rho\,\phi(0) c_{D}(uv)^{1+D/2}\frac{2D}{24(D+2)(D+1)}R(0)\right]e^{-\rho V_{0}},
\end{align}
where repeated indices are summed over and we have used $$V_{0}=C_{D}\tau^{D}=c_{D}(uv)^{D/2}.$$ 
For later convenience note that
\be\label{pt}
\omega_{D-1} = \frac{\pi^{(D-1)/2}}{\Gamma\left(\frac{D-1}{2}+1\right)}.
\ee


\section{Universality of $-R/2$ factor:}\label{III}
In order to prove the universality of the $-R/2$ factor we will construct operators $\hat{O}$ such that (see eq.~\eqref{cont2}) 
\begin{equation}\label{a}
\lim_{\rho \to +\infty}\bar{B}^{(D)}_{\rho}\phi=\Box_{g}\phi+a R\phi
\end{equation}
and then prove that $a=-1/2$ for the entire family of GCD (in every dimensions).

We firstly construct operators $\hat{O}$ that annihilate terms in eq.~\eqref{I} that would give rise to divergences in the local limit. In this way, we are going to recover eqs.~\eqref{eq:even1} and~\eqref{eq:odd1}. Then we will choose $\hat{O}$ such that we recover the d'Alembertian in the local limit and prove that this implies $a=-1/2$ in eq.~\eqref{a} for the entire family of GCD.
 
Let us start by noting that, all the integrals appearing in eq.~\eqref{I} (for $D>2$) are of the general form 
\be\label{ge}
I_{m,n}=\int_{0}^{\ta}dv\int_{0}^{v}du \,u^{m}v^{n}e^{-\rho\,c_{D}(u\,v)^{D/2}}, 
\ee
with non-negative  $m,n$ and $m+n=D-2,\,D-1,\, D,\, 2D$. This can be recognized to be true by a binomial expansion of the term $(u-v)^{D-2}$ in eq.~\eqref{I}.
 
In even dimensions we show that, eq.~\eqref{ge} with $m\neq n$ gives divergent contributions in the local limit. 
Constructing the operators $\hat{O}$ such that they annihilate these divergences ensures that they also annihilate divergent terms coming from eq.~\eqref{ge} with $m=n$. However when $m=n$ logarithmic terms, which are not annihilated by $\hat{O}$, are also present and they lead to the finite contributions in the local limit.

In odd dimensions we have $m\neq n$ always. We therefore construct $\hat{O}$ such that it annihilates the divergent terms with the exception of those needed for eliminating the first term appearing on the RHS of eq.~\eqref{cont2}, i.e. the terms with $m+n=D-2$, see eq.~\eqref{I}.

We discuss first the case of even dimensions with $D>2$ in detail and then briefly analyze the 2D as well as odd dimensional cases.
\subsection{Even dimensions: case $m\neq n$}
Consider $I_{m,n}$ for $m\neq n$,
\begin{align}\label{Inm}
I_{m\neq n}&=\frac{\ta^{-m}}{(m-n)D/2}\frac{1}{(c_{D}\rho)^{\frac{2+2m+n}{D}}}\left[\ta^{n}(c_{D}\rho)^{\frac{n}{D}}\left(-\Gamma(\frac{1+m}{D/2})+\Gamma(\frac{1+m}{D/2},\ta^{D}c_{D}\rho)\right)\right.\\ \nn
&\left.+\ta^{m}(c_{D}\rho)^{\frac{m}{D}}\left(\Gamma(\frac{2+m+n}{D})-\Gamma(\frac{2+m+n}{D},\ta^{D}c_{D}\rho)\right)\right],
\end{align} 
where $\Gamma(\cdot,\cdot)$ are incomplete Gamma functions which are exponentially vanishing in the local limit. The terms potentially divergent in the local limit are of the form $$\rho^{-\frac{2+2m}{D}}, \ \ \ \rho^{-\frac{2+m+n}{D}},$$ for $m+n\neq 2D$ and $$\rho^{-\frac{2+2m}{D}+1}, \ \ \ \rho^{-\frac{2+m+n}{D}+1},$$ for $m+n=2D$. Note that, terms with $m+n=2D$ are multiplied by $\rho$ in eq.~\eqref{I} and $m$ is always at least $D/2$ thanks to the presence of the $(uv)^{D/2}$ factor.
  
Given the factor $\rho^{\frac{2+D}{D}}$ in eq.~\eqref{cont2}, a sensible requirement for the operator $\hat{O}$ is to annihilate terms proportional to $\rho^{-\alpha}$ with $$\frac{2+D}{D}-\alpha\geq 0.$$ Indeed those are the terms that diverge for $\rho\rightarrow\infty$, i.e. in the local limit. The need for annihilate $\rho^{(D+2)/D}$ will be fully clarified in the next section.
We consider the two relevant subcases, $m+n\neq D-1$ and $m+n=D-1$, separately since the second one shows why we require $\hat{O}$ to annihilate also $\rho^{(D+2)/D}$, which in principle should give a finite contribution in the local limit.
\subsubsection{Case $m+n\neq D-1$}\label{secreq}
The relevant terms in $I_{m,n}$ can be written as $\rho^{-p/D}$ with $p$ a positive and even integer. We require that $\hat{O}$ annihilates terms for which   
\be\label{rr}
\frac{2+D}{D}-\frac{p}{D}\geq 0\Rightarrow\: p\leq 2+D,
\ee 
i.e. 
\be\label{req}
\hat{O}\rho^{-\frac{2(k+1)}{D}}=0,\;\;k=0,1,2,\dots,\frac{D}{2}.
\ee
These are $(D+2)/2$ requirements, i.e. $N+2$ requirements (defining $D=2N+2$ as in sec.\ref{even} and~\ref{odd}), exactly as many as in eq.~\eqref{eq:even1}.
Using 
\be
H_{n}\left(\frac{1}{\rho^{\frac{2(k+1)}{D}}}\right)=\frac{1}{\rho^{\frac{2(k+1)}{D}}}(-1)^{n}\prod_{s=0}^{n-1}(\frac{2(k+1)}{D}+s)=(-1)^{n}\frac{1}{\rho^{\frac{2(k+1)}{D}}}\frac{2(k+1)}{D}\frac{\Gamma[\frac{2(k+1)}{D}+n]}{\Gamma[1+\frac{2(k+1)}{D}]}.
\ee
one can show that (see eq.~\eqref{ohat})
\begin{align}\label{eq21}
\hat{O}\rho^{-\frac{2(k+1)}{D}}& =\sum_{n=0}^{L_{max}}\frac{b_{n}}{n!}\frac{2(k+1)}{D}\frac{\Gamma[\frac{2(k+1)}{D}+n]}{\Gamma[1+\frac{2(k+1)}{D}]}\\ \nn
&=\frac{2(k+1)}{D}\frac{1}{\Gamma[1+\frac{2(k+1)}{D}]}\sum_{n=0}^{L_{max}}\frac{b_{n}}{n!}\Gamma[\frac{2(k+1)}{D}+n],
\end{align}
which proves the equivalence of eqs.~\eqref{req} and~\eqref{eq:even1}.
\subsubsection{$m+n= D-1$}\label{dmenouno}
Consider the term 
\be\label{te}
\frac{\phi_{,0}(0)}{\sqrt{2}}\left(-\int_{0}^{\ta}dv\int_{0}^{v}du\frac{(v-u)^{D-2}}{2^{(D-2)/2}}(u+v)e^{-\rho'(u\,v)^{D/2}}\right),
\ee
where $\rho'\equiv c_{D}\rho$ and we now define $D=2M$. Using the change of variables
\begin{align}\label{cv}
x&\equiv u^{M},\\ \nonumber 
y&\equiv v^{M},
\end{align}
and the binomial expansion, the general integrals appearing in eq.~\eqref{te} are ($M>1$ since $D>2$)
\begin{align}\label{uno}
&\int_{0}^{\ta^{M}}dy\int_{0}^{y}dx \,y^{\frac{M-1-k}{M}}x^{\frac{2+k-M}{M}}=\\ \nn
& \frac{M\rho'^{-1-\frac{1}{2M}}\Gamma\left(1+\frac{1}{2M}\right)}{3+2k-2M}-\frac{\ta^{2M-3-2k}M\rho'^{-\frac{2+k}{M}}\Gamma\left(\frac{2+k}{M}\right)}{3+2k-2M}\\ \nn
&-\frac{M\rho'^{-1-\frac{1}{2M}}\Gamma\left(1+\frac{1}{2M},\ta^{2M}\rho'\right)}{3+2k-2M}+\frac{\ta^{2M-3-2k}M\rho'^{-\frac{2+k}{M}}\Gamma\left(\frac{2+k}{M},\ta^{2M}\rho'\right)}{3+2k-2M},
\end{align}

\begin{align}\label{due}
&\int_{0}^{\ta^{M}}dy\int_{0}^{y}dx\, y^{\frac{M-k}{M}}x^{\frac{1+k-M}{M}}=\\ \nn
& \frac{M\rho'^{-1-\frac{1}{2M}}\Gamma\left(1+\frac{1}{2M}\right)}{1+2k-2M}-\frac{\ta^{2M-1-2k}M\rho'^{-\frac{1+k}{M}}\Gamma\left(\frac{1+k}{M}\right)}{1+2k-2M}\\ \nn
&-\frac{M\rho'^{-1-\frac{1}{2M}}\Gamma\left(1+\frac{1}{2M},\ta^{2M}\rho'\right)}{1+2k-2M}+\frac{\ta^{2M-1-2k}M\rho'^{-\frac{1+k}{M}}\Gamma\left(\frac{1+k}{M},\ta^{2M}\rho'\right)}{3+2k-2M}.
\end{align}
Note that since $k$ and $M$ are integers the denominators never vanish. The incomplete Gamma functions do not contribute in the local limit, thus the relevant terms are $$\rho^{-\frac{D+1}{D}},\ \ \ \rho^{-\frac{2+k}{M}},$$ from eq.~\eqref{uno} and $$\rho^{-\frac{D+1}{D}},\ \ \ \rho^{-\frac{1+k}{M}},$$ from eq.~\eqref{due}.
To summarize:
\begin{itemize}
	\item Terms we want to get rid of that are proportional to $\rho^{-\frac{2+k}{M}}=\rho^{-\frac{4+2k}{D}}$ give a divergent or a constant term in the local limit. These terms are such that $$\frac{D+2}{D}-\frac{4+2k}{D}\geq 0\Rightarrow k\leq M-1,$$ i.e. they are of the form $\rho^{-p/D}$ with $p$ even and less than or equal to $D+2$ (see eq.~\eqref{rr}), as such they are already annihilated by $\hat{O}$, see eqs.~\eqref{rr} and~\eqref{req}. 
 \item Terms proportional to $\rho^{-\frac{D+1}{D}}$ sum to zero
	\be
	-\rho^{-\frac{D+1}{D}}\frac{D}{2}\Gamma(1+\frac{1}{D})\sum_{k=0}^{D-2}\binom{D-2}{k}(-1)^{k}\frac{4+4k-2D}{(3+2k-D)(1+2k-D)}=0. 
	\ee
\end{itemize}
We see that $\hat{O}\rho^{-(D+2)/D}=0$ comes from requiring the correct IR behavior of the operator, in the sense of obtaining $\Box$ rather than  some other combination of derivatives.    
\subsection{Even dimensions: case $m=n$}\label{mn}
Let us consider the terms with $m=n$. Note that $m$ can assume the values $\frac{D-2}{2},\frac{D}{2},D$, where the $I_{D,D}$ are also multiplied by a factor of $\rho$. The terms of interest in eq.~\eqref{I} are given by
\begin{align}\label{terms}
&\frac{1}{2^{\frac{D-2}{2}}}\left\{ (D-1)\omega_{D-1}\phi(0)A_{0}I_{\frac{D-2}{2},\frac{D-2}{2}}\right.\\ \nn
&\left.+A_{1}\left[\frac{1}{2}\omega_{D-1}\left(\frac{D-1}{2}\phi_{,00}(0)-\frac{1}{6}\phi(0)(D-1)R_{00}\right)\right]I_{D/2,D/2}\right.\\ \nn
& \left.+A_{2}\left[\frac{1}{2}\omega_{D-1}\left(\frac{1}{2}\phi_{,ii}-\frac{1}{6}\phi(0)R_{ii}\right)\right]I_{D/2,D/2}\right.\\ \nn
& \left.+A_{1}\left[\frac{\omega_{D-1}}{2}\left(-\frac{D}{24(D+1)}c_{D}\phi(0)(D-1)R_{00}\right)\right]\rho I_{D,D}\right.\\ \nn
& \left.+A_{2}\left[\frac{\omega_{D-1}}{2}\left(-\frac{D}{24(D+1)}c_{D}\phi(0)R_{ii}\right)\right]\rho I_{D,D}\right.\\ \nn
& \left.+A_{5}\left[\frac{2D}{24(D+1)(D+2)}\omega_{D-1}(D-1)c_{D}\phi(0)R\right]\rho I_{D,D}\right\},
\end{align}
where 
\begin{align}
A_{0}&\equiv\binom{D-2}{\frac{D-2}{2}}(-1)^{\frac{D-2}{2}},\\
A_{1}&\equiv\left[\binom{D-2}{\frac{D-4}{2}}(-1)^{\frac{D-4}{2}}+\binom{D-2}{\frac{D-2}{2}}2(-1)^{\frac{D-2}{2}}+\binom{D-2}{\frac{D}{2}}(-1)^{\frac{D}{2}}\right],\\
A_{2}&\equiv\left[\binom{D-2}{\frac{D-4}{2}}(-1)^{\frac{D-4}{2}}-\binom{D-2}{\frac{D-2}{2}}2(-1)^{\frac{D-2}{2}}+\binom{D-2}{\frac{D}{2}}(-1)^{\frac{D}{2}}\right],\\
A_{5}&\equiv\binom{D-2}{\frac{D-2}{2}}(-1)^{\frac{D-2}{2}}.
\end{align}
The general $I_{m,m}$ can be computed with the change of variables in eq.~\eqref{cv},
\begin{align}
\label{eq:I1} I_{D/2,D/2}&=\frac{2}{D^{2}}\rho'^{-\frac{D+2}{D}} \left[G_{2,3}^{3,0}\left(\ta^D \rho' |
\begin{array}{c}
 1,1 \\
 0,0,1+\frac{2}{D} \\
\end{array}
\right)\right.\\ \nn
&\left.+\Gamma \left(1+\frac{2}{D}\right) \left( \log (\ta^{D}\rho')-\psi\left(1+\frac{2}{D}\right)\right)\right], \\ 
I_{\frac{D-2}{2},\frac{D-2}{2}}&=\frac{4}{D^{2}}\frac{\Gamma \left(0,\ta^{D} \rho' \right)+\log (\ta^{D}\rho')+\gamma }{2 \rho' }, \label{eq:I2}\\ 
\label{eq:I3} I_{D,D}&=\frac{2}{D^{2}}\rho'^{-\frac{2}{D}-2} \left[G_{2,3}^{3,0}\left(\ta^{D} \rho' |
\begin{array}{c}
 1,1 \\
 0,0,2+\frac{2}{D} \\
\end{array}
\right)\right.\\ \nn
&\left.+\Gamma \left(2+\frac{2}{D}\right) \left(\log (\ta^{D}\rho')-\psi\left(2+\frac{2}{D}\right)\right)\right], 
\end{align}
where again $\rho'=c_{D}\rho$.
The only terms that give finite contributions in the local limit are the logarithmic ones. \footnote{Actually, the logarithmic term in eq.~\eqref{eq:I2} serves the purpose of eliminating the constant term appearing on the RHS of eq.~\eqref{cont2}, see next section.}. Indeed, terms proportional to powers of $\rho$ are annihilated\footnote{Note that the terms in eq.~\eqref{eq:I3} are multiplied by $\rho$.} by $\hat{O}$ (see eq.~\eqref{req}), whereas terms containing the $G_{2,3}^{3,0}$ Meijer's G-function do not contribute since these functions decay exponentially fast in the local limit (see~\cite{gradshteyn2009tables}).

\subsection{First condition: eliminating the constant}\label{const}
In order to not have divergences in the local limit, we need to cancel the first term appearing on the RHS of eq.~\eqref{cont2}. From eq.~\eqref{terms}, we have to impose 
\be\label{ffi}
\rho^{\frac{2+D}{D}}\frac{1}{2^{\frac{D-2}{2}}} (D-1)\omega_{D-1}A_{0}\hat{O}\underbrace{\frac{Log[\ta^{D}c_{D}\rho]}{2(\frac{D}{2})^{2}c_{D}\rho}}_{\subset I_{\frac{D-2}{2},\frac{D-2}{2}}}=-\rho^{2/D}a.
\ee
Using expression eq.~\eqref{pt}, $D=2N+2$, $c_{D}=2^{N+1}C_{2N+2}$ and noting that $$A_{0}=(-1)^{N}\frac{(2N)!}{(N!)^{2}}$$ we can rewrite~\eqref{ffi} as
\be\label{first}
\frac{1}{2^{N}}(-1)^{N}\frac{(2N)!}{(N!)^{2}} (2N+1)\frac{2 (4 \pi )^N N!}{(2 N+1)!}\frac{1}{2(N+1)^{2}2^{N+1}C_{2N+2}}\hat{O}\frac{Log[\ta^{D}c_{D}\rho]}{\rho}=-\frac{a}{\rho}.
\ee
It can be proven that this last equation is equivalent to eq.~\eqref{eq:even2}, see appendix~\ref{IndI} for details.     
\subsection{Second condition: finding the d'Alembertian}\label{44}
We now proceed by choosing $\hat{O}$ such that we recover the d'Alembertian in the local limit and prove that this implies $a=-1/2$ in eq.~\eqref{a} for the entire family of GCD.
In order to obtain the d'Alembertian from terms involving two derivatives of the field in eq.~\eqref{I} we require that (see eq.~\eqref{terms})
\begin{equation}\label{secnw}
\lim_{\rho\rightarrow\infty}\left\{\rho^{\frac{D+2}{D}}\frac{1}{2^{\frac{D}{2}}}\omega_{D-1}\left[\frac{D-1}{2}\phi_{,00}A_{1}+\frac{1}{2}\phi_{,ii}A_{2}\right]\hat{O}I_{D/2,D/2}\right\}=\Box\phi.
\end{equation}
This is equivalent to considering the action of $\hat{O}$ on the logarithmic term in eq.~\eqref{eq:I1} (see discussion thereafter). 
From \mbox{$-(D-1)/2\ A_{1}=A_{2}/2$} we have 
\begin{equation}\label{mb}
\frac{1}{2^{\frac{D}{2}}}\omega_{D-1}\left[\frac{D-1}{2}\phi_{,00}A_{1}+\frac{1}{2}\phi_{,ii}A_{2}\right]=\frac{1}{2^{\frac{D-2}{2}}}\frac{1}{2}\omega_{D-1}\frac{A_{2}}{2}\Box\phi(0),
\end{equation}
and using eq.~\eqref{mb} in eq.~\eqref{secnw} we obtain
\begin{equation}\label{secondre}
\rho^{\frac{D+2}{D}}\hat{O}\frac{\Gamma[\frac{D+2}{D}]}{2(\frac{D}{2})^{2}(c_{D})^{\frac{D+2}{D}}\rho^{\frac{D+2}{D}}}\log(\ta^{D}c_{D}\rho)=\frac{2^{\frac{D+2}{2}}}{\omega_{D-1}A_{2}}.
\end{equation}
It can be shown that this last equation is equivalent to eq.~\eqref{eq:even3}, see appendix~\ref{IndII} for details. This result should not come as a surprise since we had required to obtain the d'Alembertian in the first place. Up to now, we have bridged the gap between the formalism of~\cite{Aslanbeigi:2014tg} and the one of~\cite{Benincasa:2010ac,Dowker:2013vl, Belenchia:2015hca}. We now show that the conditions we have found imply $a=-1/2$ in eq.~\eqref{a}, i.e. that the factor $-R/2$ is universal for all GCD (in even dimensions).

\subsection{Third condition: universal factor}\label{lasteven}

Finally we must consider terms in eq.~\eqref{terms} that contain curvatures and are given by
\begin{align}\label{localR}
&\frac{1}{2^{\frac{D-2}{2}}}\left\{ -\frac{\omega_{D-1}}{12}A_{2}R\phi(0)I_{D/2,D/2}\right.\\ \nn
&\left. -\frac{\omega_{D-1}}{2}A_{2}\frac{D}{24(D+1)}c_{D}R\phi(0)\rho I_{D,D}\right.\\ \nn
& \left. +\omega_{D-1}(D-1)A_{5}\frac{2D}{24(D+1)(D+2)}c_{D}R\phi(0)\rho I_{D,D}\right\},
\end{align}
where we used $A_{1}=-A_{2}/(D-1)$. 
Note that the action of $\hat{O}$ on the first term in eq.~\eqref{localR} is completely determined by eq.~\eqref{secondre} and gives
\be\label{omam}
-\frac{R}{3}\phi.
\ee
For the other terms we need to compute $\hat{O}\rho I_{D,D}$. Since we are interested in the local limit we focus on the logarithmic term of $I_{D,D}$ (see eq.~\eqref{eq:I3} and discussion thereafter) 
\be\label{cavl}
\hat{O}\left(\frac{2}{D^{2}c_{D}^{\frac{2D+2}{D}}}\Gamma(\frac{2D+2}{D})\frac{\log(\ta^{D}\rho)}{\rho^{\frac{D+2}{D}}}\right).
\ee
It is easy to see that eq.~\eqref{cavl} is determined by eq.~\eqref{secondre} since
\begin{align}\label{63}
&\rho^{\frac{D+2}{D}}\hat{O}\left(\frac{2}{D^{2}}\Gamma(\frac{2D+2}{D})\frac{\log(\ta^{D}\rho)}{\rho^{\frac{D+2}{D}}}\right)\\ \nn
&=\rho^{\frac{D+2}{D}}\frac{\Gamma(\frac{2D+2}{D})}{\Gamma(\frac{D+2}{D})c_{D}}\hat{O}\left(\frac{\Gamma[\frac{D+2}{D}]}{2(\frac{D}{2})^{2}(c_{D})^{\frac{D+2}{D}}\rho^{\frac{D+2}{D}}}\log(\ta^{D}c_{D}\rho)\right)\\ \nn
&= \frac{\Gamma(\frac{2D+2}{D})}{\Gamma(\frac{D+2}{D})c_{D}}\frac{2^{\frac{D+2}{2}}}{\omega_{D-1}A_{2}}.
\end{align}
Using eq.~\eqref{63} we find that the last two terms in eq.~\eqref{localR} give in the local limit 
\begin{align}\label{uff}
&-\frac{1}{6}\frac{D+2}{2D+2}R\phi(0)\\ 
&\frac{D-1}{D+1}\frac{1}{3}\frac{A_{5}}{A_{2}}R\phi(0),
\end{align}
where it can be shown that $A_{5}/A_{2}=D/(4-4D)$. Summing eq.~\eqref{omam} and eq.~\eqref{uff} we finally obtain the universal factor 
\begin{equation}
\boxed{
\left(-\frac{1}{3}-\frac{1}{6}\frac{D+2}{2D+2}-\frac{D}{12 (D+1)}\right)R(0)\phi(0)=-\frac{1}{2}R(0)\phi(0).
}
\end{equation}
Hence, we have proven that all GCD in even dimensions reduce to \mbox{$(\Box-R/2)\phi$} in the local limit.


\subsection{2D case:}
The only difference with the previous sections is that $r$ is no more a non-negative radial coordinate and eq.~\eqref{I} is replaced by 
\begin{align}
I(2)=&\int_{0}^{\tilde{a}}dv\int_{0}^{\tilde{a}}du\left[\phi(0)+r \phi_{,r}+\frac{-u-v}{\sqrt{2}}\phi_{,0}\right. \\ \nn
&\left.+\frac{1}{2}\frac{(u+v)^{2}}{2}\phi_{,00}+\frac{1}{2}\frac{(v-u)^{2}}{2}\phi_{,rr}+ r\frac{(v-u)}{\sqrt{2}}\phi_{,0r}\right.\\ \nn
&\left.-\frac{1}{6}R_{rr}\phi \frac{(v-u)^{2}}{2}-\frac{1}{6}R_{00}\phi \frac{(v+u)^{2}}{2}-\frac{1}{6}R_{0r}\phi 2r \frac{-u-v}{\sqrt{2}}\right.\\ \nn
&\left.-\rho\phi uv\left(-\frac{R}{72}uv+\frac{R_{00}}{36}\frac{(v+u)^{2}}{2}\right) \right]e^{-\rho u v}.
\end{align}
With this clarification, it is possible to proceed in the analysis as in the previous sections. In particular, requiring $\hat{O}$ to annihilate the diverging terms (in the local limit) we arrive at eq.~\eqref{req} for $k=0,1$. As in sec.\ref{dmenouno}, the requirement of annihilating $\rho^{-2}$ comes from eliminating the term with $\phi_{,0}$. 

When $m=n$ the only relevant terms in the local limit are again the logarithmic ones, analogously to sec.\ref{mn}. From $I_{0,0}$ the logarithmic term is given by $\rho^{-1}\log(\tilde{a}^2\rho)$. The condition for eliminating the first term in eq.~\eqref{cont2} can be obtained following sec.\ref{const}, and proved to be equivalent to eq.~\eqref{eq:even2}. Considering the integrals $I_{1,1}$ and $\rho I_{2,2}$, the only relevant term is $\rho^{-2}\log(\tilde{a}^2\rho)$ in both cases and the same calculations of sec.~\ref{44} can be applied, which completes the proof.

\subsection{Odd dimensions}\label{oddy}
In this case $m+n=D-2,\:D,\:2D$ (the case with $n+m=D-1$ will be considered separately) in eq.~\eqref{ge} and $m\neq n$ always (see appendix~\ref{app:extratsub} for a simple proof of this statement), thus we refer to eq.~\eqref{Inm}. 
The terms in~\eqref{I} that are supposed to give the d'Alembertian (and the term proportional to the field which is needed to cancel the first term in eq.~\eqref{cont2}) have $m\neq n$, with $m,n$ both integers and with $m+n=D-2,D$. As in even dimensions, we require $\hat{\mathcal{O}}$ to annihilate terms that give divergences in the local limit. Considering terms proportional to $\rho^{-(2+2m)/D}$ we require that 
\begin{equation}\label{odd7}
\hat{\mathcal{O}}\rho^{-\frac{2(k+1)}{D}}=0\ \ \ k=0,1,\dots,\frac{D-1}{2}.
\end{equation}
These are $(D+1)/2$ relations, as many as the equations in eq.~\eqref{eq:odd1}. Indeed, eqs.~\eqref{eq:odd1} and~\eqref{odd7} are equivalent, as can be proved using eq.~\eqref{eq21} with $D=2N+1$. 
Terms proportional to $\rho^{-(2+m+n)/D}$ (for $m+n=D-2,D$) are potentially problematic. It is tempting to require $\hat{O}$ to also annihilate these, however in this case this requirement is too restrictive since it would annihilate all terms that can give the d'Alembertian (and the term proportional to the field) in the local limit. This is in contrast to the even dimensional case where there are terms with $m=n$.

For $m+n=2D$ we see from eq.~\eqref{I} that $m$ (and $n$) are of the form \mbox{\textit{integer}+$D/2$} and the terms are multiplied by $\rho$. Thus, problematic terms that are proportional to $\rho^{-(2+2m-D)/D}$ are annihilated by $\hat{O}$ due to eq.~\eqref{odd7}. Regarding terms proportional to $\rho^{-(2+D)/D},$ we have already argued that requiring $\hat{O}$ to annihilate them would preclude the possibility to recover the d'Alembertian in the local limit.

\subsubsection{Case $m+n=D-1$}
The term of interest is the one with the single time-derivative of the field which does not give any finite contribution in the local limit. Indeed, using the result of sec.~\ref{dmenouno} it is possible to show that all the potentially problematic terms are annihilated by $\hat{O}$ due to eq.~\eqref{odd7}.  


\subsubsection{First condition: eliminating the constant}
The first term in eq.~\eqref{cont2} has to be canceled by the one proportional to the field in eq.~\eqref{I}, i.e.
\begin{equation}\label{perconst}
 \rho^{\frac{2+D}{D}}\frac{1}{2^{\frac{D-2}{2}}}(D-1)\omega_{D-1}\hat{\mathcal{O}}\left(\int_{0}^{\tilde{a}}dv\int_{0}^{v}du(v-u)^{D-2}e^{-\rho c_{D}(uv)^{D/2}}\right)=-\rho^{2/D}a,
\end{equation}
where equality is intended in the local limit. The generic term in the integral appearing in eq.~\eqref{perconst} can be computed with the change of variables given in eq.~\eqref{cv}, leaving
\begin{equation}
\frac{-D}{2}\frac{4}{D^{2}}\frac{1}{c_{D}\rho}\sum_{k=0}^{D-2}(-1)^{k}\frac{1}{(D-2-2k)},
\end{equation}
as the only relevant terms in the local limit. Note that the denominators  never vanish since $D$ is odd. 
Calling $D=2N+1$ and using $$\sum_{k=0}^{D-2}(-1)^{k}\frac{1}{(D-2-2k)}=\frac{(-1)^{N+1}2^{-2+2N}\sqrt{\pi}(N-1)!}{\left(N-\frac{1}{2}\right)!},$$ it is easy to see that eq.~\eqref{perconst} is equivalent to eq.~\eqref{eq:odd2}.
\subsubsection{Finding the local limit}\label{lastodd}

Let us proceed as in the even dimensional case and require that the operator gives the d'Alembertian when acting on the terms containing two derivatives of the field, in the local limit, i.e. 
\begin{align}\label{boxr}
& \rho^{\frac{2+D}{D}}\hat{\mathcal{O}}\left(\int_{0}^{\ta}dv\int_{0}^{v}du\frac{(v-u)^{D-2}}{2^{(D-2)/2}} \left[\frac{(u+v)^{2}}{2}\omega_{D-1}(D-1)\frac{1}{2}\phi_{,00}(0)\right.\right.\\ \nonumber
&\left.\left.+\frac{(v-u)^{2}}{2}\omega_{D-1}\frac{1}{2}\phi_{,ii}(0)\right]e^{-\rho c_{D}(uv)^{D/2}}\right)=\Box\phi.
\end{align} 
The terms non-vanishing in the local limit, coming from eq.~\eqref{boxr} are
\begin{align}
&-\frac{8\Gamma\left(2/D\right)}{D^{2}}\left[\sum_{k=0}^{D-2}(-1)^{k}\binom{D-2}{k} \frac{2(2-4D+D^2 +8k-4D k+4k^2 )}{(D-4-2k)(D-2k)(D-2-2k)}\right]\rho^{-\frac{D+2}{D}},\\ \nonumber
&-\frac{8\Gamma\left(2/D\right)}{D^{2}}\left[\sum_{k=0}^{D}(-1)^{k}\binom{D}{k}\frac{1}{2(D-2k)}\right]\rho^{-\frac{D+2}{D}},
\end{align}
from the first and the second term in eq.~\eqref{boxr} respectively. Given that 
\be\label{lab}
\hat{\mathcal{O}}\rho^{-\frac{D+2}{D}}=\sum_{n=0}^{L_{max}}\frac{b_{n}}{n!}\frac{\Gamma\left(n+\frac{2N+3}{2N+1}\right)}{\Gamma\left(\frac{2N+3}{2N+1}\right)}\rho^{-\frac{D+2}{D}},
\ee
where we used $\Gamma(1+x)=x\Gamma(x)$, and using  
\begin{align}
& (D-1)\sum_{k=0}^{D-2}(-1)^{k}\binom{D-2}{k} \frac{2(2-4D+D^2 +8k-4D k+4k^2 )}{(D-4-2k)(D-2k)(D-2-2k)}\\ \nonumber
&= \sum_{k=0}^{D}(-1)^{k}\binom{D}{k}\frac{1}{2(D-2k)},
\end{align}
it can be seen that eq.~\eqref{boxr} is equivalent to eq.~\eqref{eq:odd3}. As in even dimensions, this result should come as no surprise. 
\subsubsection{Universality of $-R/2$ factor}
Finally, we consider the action of $\hat{O}$ on terms involving curvatures in eq.~\eqref{I} and show that $a=-1/2$ in eq.~\eqref{a}. Consider first the terms in eq.~\eqref{I} given by
\begin{align}
&\int_{0}^{\ta}dv\int_{0}^{v}du\frac{(v-u)^{D-2}}{2^{(D-2)/2}}\left[\frac{(u+v)^{2}}{2}\omega_{D-1}(D-1)\left(-\frac{1}{6}R_{00}(0)\phi(0)\right)\right.\\ \nonumber
&\left.+\frac{(v-u)^{2}}{2}\omega_{D-1}\left(-\frac{1}{6}R_{ii}(0)\phi(0)\right)\right].
\end{align} 
Using eq.~\eqref{boxr} (or equivalently eq.~\eqref{eq:odd3}) it can be shown that these terms contribute $-R/3\ \phi$ in the local limit. The remaining terms which have not been considered so far are the ones with $m+n=2D$, i.e.
\begin{align}\label{Iod}
&\int_{0}^{\ta}dv\int_{0}^{v}du\frac{(v-u)^{D-2}}{2^{(D-2)/2}} \left[\frac{(u+v)^{2}}{2}\omega_{D-1}(D-1)(-\rho)\,\phi(0) c_{D}(uv)^{D/2}\frac{D}{24(D+1)}R_{00}(0)\right.\\ \nn
&\left.\frac{(v-u)^{2}}{2}\omega_{D-1}(-\rho)\,\phi(0) c_{D}(uv)^{D/2}\frac{D}{24(D+1)}R_{ii}(0)\right.\\ \nn
&\left.+ \omega_{D-1}(D-1)\rho\,\phi(0) c_{D}(uv)^{1+D/2}\frac{2D}{24(D+2)(D+1)}R(0)\right]e^{-\rho c_{D}(uv)^{D/2}}.
\end{align}
Using eq.~\eqref{cv} the terms giving non-vanishing contributions in the local limit, are found to be proportional to $\rho^{-\frac{D+2}{D}}$ so that we can use eq.~\eqref{lab} in eq.~\eqref{Iod}. 
The final result is that the contribution of the terms in eq.~\eqref{Iod} is dimension independent and equal to $-R/6\ \phi$. Considering both curvature contributions we find that $-R/2$ is a universal factor for all GCD in odd dimensions. 
\\
This completes the proof for GCD in all dimensions.
\section{Results and discussions}\label{V}
We have studied the class of \textit{Generalized Causal Set d'Alembertians} in curved spacetime. 
In particular, we have shown that, when a local limit exists, all GCD give$$\Box_{g}\phi(x)-\frac{1}{2}R(x)\phi(x),$$ in this limit.

In doing so we have bridged the gap between the formalism of~\cite{Benincasa:2010ac,Dowker:2013vl, Belenchia:2015hca} and the one of~\cite{Aslanbeigi:2014tg}, showing how the equations found in~\cite{Aslanbeigi:2014tg} via a spectral analysis of the non-local operators in flat spacetime translate into properties of $\hat{\mathcal{O}}$ in the set-up of~\cite{Benincasa:2010ac,Dowker:2013vl, Belenchia:2015hca}. We have also shown that the requirements that lead to the right local limit in the flat case are sufficient to ensure the appearance of the universal $-R/2$ factor in curved spacetime for all GCD. 
The present result is an independent proof of the universality of the $-R/2$ factor for the entire family of non-local operators inspired by Causal set theory. Moreover, this result shows that the universal factor is not related to the minimality condition but to the physical requirements that characterize the operators.

It should be noted that the assumptions made in this work --- in particular compact support of the field --- are ubiquitous in the literature~\cite{Benincasa:2010ac,Dowker:2013vl,Glaser:2013sf,Belenchia:2015hca}. In order to weaken them, further studies are required. Extending the analysis of~\cite{Belenchia:2015hca} to all GCD would allow one to fully take into account the case in which the support of the field is not restricted to the near region. Even better, a spectral analysis similar to that in~\cite{Aslanbeigi:2014tg} would remove the assumption of compact support altogether and, as such, deserves further investigation.  

As a future direction, it would be interesting to study the connection of the result of this work with Einstein's equivalence principle (EEP). Indeed, the EEP for a scalar field coupled to gravity requires there to be a non-minimal coupling in order to hold true. This point was discussed in~\cite{Sonego:1993fw}, where the authors proved, without relying on conformal invariance, that in 4D the required coupling has to coincide with the conformal one. Thus, the value of the universal factor for GCD has to be carefully considered in light of the EEP,\footnote{Note that the value of the universal factor cannot be interpreted \textit{a priori} as violating the EEP since other effects may come about when quantum corrections are taken into account, e.g. the running of the coupling.} also in view of possible phenomenological consequences. 

In particular, for non-conformal couplings in 4D (as in our case) wave tails propagating inside the light cone are present for massless fields~\cite{Sonego:1993fw}. This amount to violations of Hyugens' principle 
which can have interesting effects (see e.g.~\cite{Jonsson:2014lja} for information theoretic consequences and~\cite{Blasco:2015eya} for implications for early Universe cosmology). More problematic would be the case of massive fields without conformal coupling in which case massive particles could be allowed to propagate on the light cone (see~\cite{Sonego:1993fw}) in clear contradiction with the local special relativistic description of physics dictated by EEP. It would be interesting to find a Causal set version of the Klein-Gordon operator\footnote{See comments both in~\cite{Belenchia:2014fda} and~\cite{Saravani:2015rva}.} and see the curved spacetime version of this operator in the local limit. We speculate that, a massive operator arising from causal set will have a local limit in curved spacetime with a curvature term different from the one of the massless case studied in this work. Whether the new term would (or could) be compatible with EEP, therefore avoiding the problematic propagation along the null cone, is entirely an open question. 

Finally, since the family of GCD is derived from a set of precise physical assumptions (see sec.\ref{GCDu}) it is tempting to understand which of these need to be relaxed/modified in order to obtain a different local limit and maybe recover the conformal coupling in 4D. We hope to come back to these points in future publications.  

\section*{Acknowledgments}
The author would like to thank Dionigi Benincasa and Stefano Liberati for helpful comments during early stages of the project and on earlier drafts. 
Special thanks also go to Eolo Di Casola and Fay Dowker for stimulating discussion and to the participants to the conference \textit{Prospects for causal set quantum gravity} held at ICMS in Edinburgh this year.
This publication was made possible through the support of the grant from the John Templeton 
Foundation No.51876. The opinions expressed in this publication are those of the authors and do not 
necessarily reflect the views of the John Templeton Foundation.




      \bibliographystyle{unsrt}
      \bibliography{mylibrary}
			
\begin{appendices}

\section{}\label{app:extrat}
In this appendix we show for completeness that the terms in eq.~\eqref{allt} which were neglected in the main text due to dimensional arguments, are indeed irrelevant in the local limit. Terms containing unknown functions (like the ones that appear in the expansion of field, metric and volume) and also the infinite series are not fully taken into account in this way, but this goes beyond the scope of the present work. However, we see no reasons why the results in~\cite{Belenchia:2015hca} should not extend to all dimensions and for non-minimal operators, given that they rely only on properties of $\hat{O}$. 

Let us start by considering the terms in eq.~\eqref{allt} given by 
\begin{align}
&\int_{0}^{\ta}dv\int_{0}^{v}du\frac{(v-u)^{D-2}}{2^{(D-2)/2}}\int d\Omega_{D-2} \\
&\left[-\frac{1}{6}R_{\mu\nu}(0)y^{\mu}y^{\nu}\cdot(y^{\mu}\phi_{,\mu}(0)+\frac{1}{2}y^{\nu}y^{\mu}\phi_{,\nu\mu}(0))\right.\\ \nn
&\left.+\left(y^{\mu}\phi_{,\mu}(0)+\frac{1}{2}y^{\nu}y^{\mu}\phi_{,\nu\mu}(0)\right)(-\rho\delta V)\right]e^{-\rho V_{0}},
\end{align}
Using spherical symmetry and with a schematic way of writing we can classify these terms based on powers of $u$ and $v$ (up to the common exponential factor) as   
\begin{equation}\label{3e4}
(v-u)^{D-2}\cdot
\begin{cases} 
t r^2, t^3  \\ 
\rho \tau^{D+2} r^2 \\
\rho \tau^{D+2} r^2 t, \rho \tau^{D+2} t^3\\
\rho \tau^{D+2} r^4, \rho \tau^{D+2} r^2 t^2, \rho \tau^{D+2} t^4\\
\rho \tau^{D} r^6,\rho \tau^{D} r^4 t^2, \rho \tau^{D} t^6
\end{cases}
\end{equation}
The general term is of the usual form $I_{m,n}$ with $m+n=\left(D+1,2D+2,2D+3,2D+4\right)$. 
The remaining terms in eq.~\eqref{allt} are given by
\begin{align}\label{lasttwo}
&\int_{0}^{\ta}dv\int_{0}^{v}du\frac{(v-u)^{D-2}}{2^{(D-2)/2}}\int d\Omega_{D-2} \\
& \left[\left(\phi(0)+y^{\mu}\phi_{,\mu}(0)+\frac{1}{2}y^{\nu}y^{\mu}\phi_{,\nu\mu}(0)\right)\delta\sqrt{-g}\cdot(-\rho\delta V)\right.\\ \nn
& \left.+(1+\delta\sqrt{-g})\cdot (\phi(y))\cdot \sum_{k=2}^{\infty}\frac{(-\rho\delta V)^{k}}{k!}\right]e^{-\rho V_{0}}.
\end{align}
In schematic form
\begin{equation}\label{5}
(v-u)^{D-2}\cdot
\begin{cases} 
\rho \tau^{D+2} r^2, \rho \tau^{D+2} t^2  \\ 
\rho \tau^{D+2} r^2 t, \rho \tau^{D+2} t^3\\
\rho \tau^{D+2} r^4, \rho \tau^{D+2} r^2 t^2, \rho \tau^{D+2} t^4\\
\rho \tau^{D} r^4,\rho \tau^{D} r^2 t^2, \rho \tau^{D} t^4 \\
\rho \tau^{D} r^4 t,\rho \tau^{D} r^2 t^3, \rho \tau^{D} t^5 \\
\rho \tau^{D} r^6,\rho \tau^{D} r^2 t^4, \rho \tau^{D} r^4 t^2, \rho \tau^{D} t^6 ,
\end{cases}
\end{equation}
for terms in the first line of eq.~\eqref{lasttwo} and 
\begin{equation}\label{6}
(v-u)^{D-2}\cdot
\begin{cases} 
\rho^k \tau^{Dk+2k}, \rho^k \tau^{Dk} y^2k  \\ 
\rho^k \tau^{Dk+2k} y, \rho^k \tau^{Dk}y^{2k+1}\\
\rho^k \tau^{Dk+2k} y^2, \rho^k \tau^{Dk}y^{2k+2}\\
\rho^k \tau^{Dk+2k} y^3, \rho^k \tau^{Dk}y^{2k+3}\\
\rho^k \tau^{Dk+2k} y^4, \rho^k \tau^{Dk}y^{2k+4},
\end{cases}
\end{equation}
for terms in the second line, where $k\geq 2$ and $y$ can be both $r$ and $t$ (with $r$ always appearing in even powers due to spherical symmetry). Terms in eqs.~\eqref{5} and~\eqref{6} contain $I_{m,n}$ with 
$$m+n=
\begin{cases}
2D+2,2D+3,2D+4,(k+1)D+2(k-1)\\
(k+1)D+2k-1,(k+1)D+2k,(k+1)D+2k+1,(k+1)D+2(k+1)).
\end{cases}$$ 

Now that we have collected all the terms of interest we can study their contributions in the local limit. We separate the discussion into even and odd dimensions. 
\subsection{Even dimensions}
The general terms that appear in the previous section are of the form $I_{m,n}$ (multiplied by some power of $\rho$). We need to differentiate two cases.
\subsubsection{Case $n\neq m$}
The relevant terms in the local limit are proportional to (see eq.~\eqref{secreq}) 
$$\rho^{-\frac{2+2m}{D}}, \rho^{-\frac{2+m+n}{D}}$$ if $m+n=D+1$, $$\rho^{-\frac{2+2m-D}{D}}, \rho^{-\frac{2+m+n-D}{D}}$$ for other values of $m+n$ not involving $k\geq 2$ and $$\rho^{-\frac{2+2m-D k}{D}}, \rho^{-\frac{2+m+n-D k}{D}}$$ for the terms in eq.~\eqref{6}.
Given the definition of $\hat{O}$ (see eq.~\eqref{ohat}) only terms proportional to $\rho^{-\alpha}$ with $\alpha\leq\frac{D+2}{D}$ can give divergent or finite contributions in the local limit. 
\begin{itemize}
	\item  For $m+n=D+1$ the only terms that could give problems are the ones proportional to $\rho^{-\frac{2+2m}{D}}$, but these are annihilated by $\hat{O}$, see eq.~\eqref{req}. 
	\item For all the other terms not involving $k$, there is always a factor of $\tau^{D}=(uv)^{D/2}$, i.e. $m$ (or $n$) is always of the form $D/2+x$ with $x$ an integer. Possible divergent (or finite) terms are proportional to $\rho^{-\frac{2+2m-D}{D}}=\rho^{-\frac{2+2x}{D}}$ and are annihilated by $\hat{O}$. 
		\item Finally, for the terms in eq.~\eqref{6}: the ones proportional to $\rho^{-\frac{2+m+n-D k}{D}}$ do not give any contribution in the local limit; the ones proportional to $\rho^{-\frac{2+2m-D k}{D}}$ are annihilated by $\hat{O}$ since there is always a factor $\tau^{D k}$, i.e. $m=kD/2+x$ with $x$ integer. 
\end{itemize}

\subsubsection{Case $m=n$}
In this case all the terms are multiplied by $\rho$ or $\rho^{k}$ and it is easy to see (by directly computing the integrals using eq.~\eqref{cv}) that the relevant terms are $$\rho^{\frac{D-2-2m}{D}},\rho^{\frac{D k-2-2m}{D}}$$ and  $$\rho^{\frac{D-2-2m}{D}} \log(\tilde{a}^{D}\rho),\rho^{\frac{D k-2-2m}{D}}\log(\tilde{a}^{D}\rho).$$
It can be shown that $$\hat{O}\rho^{-\alpha}\log(c \rho)\propto \rho^{-\alpha}\log(c \rho),$$ thus these terms do not give any finite contribution in the local limit.

\subsection{Odd dimensions}\label{app:extratsub}
In odd dimensions the general term is of the form $I_{m,n}$ with $m\neq n$. To see this, consider the general form of the integrands of the terms of interest
\begin{align}
(v-u)^{D-2} \tau^{C D+2C} r^{2A} t^{B}&\approx  (v-u)^{D-2} (uv)^{\frac{C D}{2}+C} (v-u)^{2A} (u+v)^{B}\\ \nn
& =(uv)^{\frac{C D}{2}+C} (v-u)^{X} (u+v)^{B},  
\end{align}
where $A,B,C$ are non-negative integers and $X=D-2+2A$ is an odd and positive integer since we consider always $D\geq 3$. Is there a monomial, in the above expression, of the form $u^{m}v^{m}$? The answer is no, as one can see by expanding the last two terms and using the structure of the binomial coefficients.

The  relevant terms in $I_{m,n}$, in the local limit, are given by 
\be\label{finay}
\rho^{-\frac{2+m+n}{D}},\rho^{-\frac{2+2m}{D}};\rho^{-\frac{2+m+n-D}{D}},\rho^{-\frac{2+2m-D}{D}}; \rho^{-\frac{2+m+n-D k}{D}}, \rho^{-\frac{2+2m-D k}{D}},
\ee 
coming form the first term in eq.~\eqref{3e4}, the terms in eq.~\eqref{3e4} and eq.~\eqref{5} multiplied by $\rho$ and the terms in eq.~\eqref{6} respectively. Due to the values that $m+n$ can assume the terms that could give finite (or divergent) contributions in the local limit are the ones in which $m+n$ does not appear. The first term in eq.~\eqref{finay} is annihilated by $\hat{\mathcal{O}}$, see eq.~\eqref{odd7}. The other terms in eq.~\eqref{finay} comes from expressions in which are always present $\tau^{D}$ or $\tau^{D k}$ respectively, i.e. $m=\frac{j\,D}{2}+ M$ where $M$ is an integers and  $j\geq 1$ (in order to take into account both cases, with and without $k$). Thus, the terms of interest are proportional to $$\rho^{-\frac{2+2M}{D}}$$ and so are annihilated by $\hat{\mathcal{O}}$.

This concludes the treatment of the extra terms of eq.~\eqref{allt} that were neglected in the main text due to dimensional arguments.

\section{}\label{app:B}
In this appendix we collect some details of the proof that were omitted in sec.~\ref{III}.

\subsection{Equivalence of eqs.~\eqref{first} and~\eqref{eq:even2}}\label{IndI}
To prove the equivalence we need first of all to compute $\hat{O}\left(Log[\ta^{D}c_{D}\rho]/\rho\right)$. From eq.~\eqref{ohat} we see that it is sufficient to compute the general 
\be
H_{n} \frac{Log[c\rho]}{\rho},
\ee
where $c$ is a constant:
\begin{align}\label{formn}
H_{n} \frac{Log[c\rho]}{\rho}&=\frac{1}{\rho}\left(A_{n}^{1}+A_{n}^{2}\log[c\rho]\right),\\
A_{n}^{1}&=(-1)^{n+1}n!(\psi(n+1)+\gamma),\\ \nonumber
A_{n}^{2}&=(-1)^{n}n!. \nonumber
\end{align}
We prove eq.~\eqref{formn} by induction. Assuming that the $n$-th term is of the form in eq.~\eqref{formn}, we want to prove that the $n+1$-th term is of the same form. From the definition of $H_{n}$ 
\be
H_{n+1}(\cdot)\equiv\rho^{n+1}\frac{\partial}{\partial\rho^{n+1}}(\cdot)=-n\,H_{n}(\cdot)+H(H_{n}(\cdot)),
\ee
so that 
\begin{align}
H_{n+1} \frac{Log[c\rho]}{\rho}&=-\frac{n}{\rho}\left(A_{n}^{1}+A_{n}^{2}\log[c\rho]\right)\\ \nn
&+A^{n}_{1}\left(-\frac{1}{\rho}\right)+A^{n}_{2}\frac{1-\log[c\,\rho]}{\rho}\\ \nn
&=\frac{1}{\rho}\left[\left(-nA^{n}_{1}-A^{n}_{1}+A^{n}_{2}\right)+\left(-nA^{n}_{2}-A^{n}_{2}\right)\log[c\rho]\right].
\end{align}
Consider the two new coefficients in the above expression, the first one is 
\begin{align}
\left(-nA^{n}_{1}-A^{n}_{1}+A^{n}_{2}\right)&=\\ \nn
&=(-1)^{n+1+1}(n+1)!(\psi(n+1)+\gamma+\frac{n!}{(n+1)!})\\ \nn
&=(-1)^{n+1+1}(n+1)!(\psi(n+1+1)+\gamma)\equiv A^{n+1}_{1},
\end{align}
where in the last line we used $\psi(x+1)=\psi(x)+\frac{1}{x}$. The second coefficient is  
\begin{align}
\left(-nA^{n}_{2}-A^{n}_{2}\right)&=\\ \nn
&=-(-1)^{n}(n+1)n!=(-1)^{n+1}(n+1)!\equiv A^{n}_{2}.
\end{align}
This conclude the inductive proof of eq.~\eqref{formn}.

Inserting eq.~\eqref{formn} in eq.~\eqref{first} we have
\begin{align}\label{quaf}
& \frac{1}{2^{N}}(-1)^{N}\frac{(2N)!}{(N!)^{2}} (2N+1)\frac{2 (4 \pi )^N N!}{(2 N+1)!}\frac{1}{2(N+1)^{2}2^{N+1}C_{2N+2}}\cdot\\ \nn
& \sum_{n=0}^{L_{max}}\frac{b_{n}}{n!}(-1)^{n}\left[(-1)^{n+1}n!(\psi(n+1)+\gamma)+(-1)^{n}n!\log[c_{D}\ta^{D}\rho]\right]=-a.
\end{align}
The sum $\sum_{n=0}^{L_{max}}b_{n}$ appearing in the above expression vanishes, see eq.~\eqref{eq:even1} with $k=\frac{D-2}{2}=N$, therefor eq.~\eqref{quaf} reduces to
\begin{equation}
a+\frac{2(-1)^{N+1}\pi^{N}}{C_{D}N!D^{2}}\sum_{n=0}^{L_{max}}b_{n}\psi(n+1)=0,
\end{equation}
i.e. eq.~\eqref{eq:even2}.

\subsection{Equivalence of eqs.~\eqref{secondre} and~\eqref{eq:even3}}\label{IndII}
To prove the equivalence we need first of all to compute $\hat{O}\left[\log(\ta^{D}c_{D}\rho)/\rho^{(D+2)/D}\right]$. We prove by induction that
\begin{align}\label{fffor}
H_{n} \frac{Log[c\rho]}{\rho^{\frac{D+2}{D}}}&=\frac{1}{\rho^{\frac{D+2}{D}}}\left(B_{n}^{1}+B_{n}^{2}\log[c\rho]\right),\\
B_{n}^{1}&=(-1)^{n+1}\frac{1}{\Gamma(\frac{D+2}{D})}\left[\Gamma(n+\frac{D+2}{D})\psi(n+\frac{D+2}{D})-\psi(\frac{D+2}{D})\Gamma(n+\frac{D+2}{D})\right],\\ \nn
B_{n}^{2}&=(-1)^{n}\frac{\Gamma(n+\frac{D+2}{D})}{\Gamma(\frac{D+2}{D})}. \nn
\end{align}
Assume that the $n$-th term is of the above form, then the $n+1$-th term is given by 
\begin{align}
H_{n+1} \frac{Log[c\rho]}{\rho^{\frac{D+2}{D}}}&=\frac{1}{\rho^{\frac{D+2}{D}}}\left[\left(-n B_{n}^{1}+B_{n}^{1}\frac{-2-D}{D}+B^{n}_{2}\right)\right.\\ \nn
&\left.+\left(-n B^{2}_{n}-\frac{D+2}{D}B^{2}_{n}\right)\log(c\rho)\right].
\end{align}
The coefficients in the above expression are such that
\begin{align}
&\left(-n B_{n}^{1}+B_{n}^{1}\frac{-2-D}{D}+B^{2}_{n}\right)\\ \nn
&= (-1)^{n+1+1}\frac{1}{\Gamma(\frac{D+2}{D})}\left[\Gamma(n+1+\frac{D+2}{D})\psi(n+1+\frac{D+2}{D})-\psi(\frac{D+2}{D})\Gamma(n+1+\frac{D+2}{D})\right]\\ \nn
&\equiv B_{n+1}^{1},
\end{align} 
\begin{align}
& \left(-n B^{2}_{n}-\frac{D+2}{D}B^{2}_{n}\right)=\\ \nn
&=(-1)^{n+1}\frac{\Gamma(n+1+\frac{D+2}{D})}{\Gamma(\frac{D+2}{D})}\equiv B^{2}_{n+1},
\end{align}
where we used $\psi(x+1)=\psi(x)+1/x$ and $\Gamma(1+x)=x\Gamma(x)$. This concludes the proof by induction of eq.~\eqref{fffor}.

We now have 
\be
\hat{O}\left[\log(\ta^{D}c_{D}\rho)/\rho^{(D+2)/D}\right]=\sum_{n=0}^{L_{max}}\frac{b_{n}}{n!}(-1)^{n}(B^{n}_{1}+B^{n}_{2}\log(\ta^{D}c_{D}\rho)),
\ee
where $\sum_{n=0}^{L_{max}}\frac{b_{n}}{n!}(-1)^{n}B^{n}_{2}=0$ (see eq.~\eqref{eq:even1} with $k=D/2$). Using this eq.~\eqref{secondre} became
\be
\sum_{n=0}^{L_{max}}\frac{b_{n}}{n!}(-1)^{n}B^{n}_{1}=\frac{2^{N+2}}{\frac{2N!(4\pi)^{N}}{(2N+1)!}}\frac{2(N+1)^{2}2^{N+2}C_{D}^{\frac{N+2}{N+1}}}{\Gamma(\frac{D+2}{D})}\frac{1}{A_{2}},
\ee
where we used the expressions for $\omega_{D-1}$ and $c_{D}$. The RHS of the above expression can be further simplified observing that
$$A_{2}=(-1)^{N+1}2\cot(2N)!\frac{\frac{2N+1}{N+1}}{(N!)^{2}}.$$ In this way we obtain for eq.~\eqref{secondre}
\be\label{simply}
\sum_{n=0}^{L_{max}}\frac{b_{n}}{n!}(-1)^{n}B^{n}_{1}=\frac{2}{\pi^{N}}C_{D}^{\frac{N+2}{N+1}}D^{2}(N+1)!\frac{(-1)^{N+1}}{\Gamma(\frac{D+2}{D})}.
\ee
Finally, using that $\sum_{n=0}^{L_{max}}\frac{b_{n}}{n!}(-1)^{n}\Gamma(n+\frac{D+2}{D})=0$ (see eq.~\eqref{eq:even1} with $k=D/2$) and the expression of $B^{n}_{1}$ from eq.~\eqref{fffor} we conclude that eq.~\eqref{secondre} is equivalent to eq.~\eqref{eq:even3}.

\end{appendices}
\end{document}